%% file: main.tex
\pdfoutput=1
\documentclass[journal,transmag]{IEEEtran}

\usepackage{paralist}
\usepackage{color}
\usepackage{siunitx}
\usepackage{url}
\usepackage{breakurl}
\usepackage{authblk} 
\usepackage[breaklinks,pdftex,
            pdfauthor={A. Bartoli, E. Medvet},
            pdftitle={Citation Counts and Evaluation of Researchers in the Internet Age},
            pdfkeywords={bibliometric, research evaluation, citations}]{hyperref}
\usepackage{tikz}
\usepackage{subcaption}

\begin{document}
\title{Citation Counts and Evaluation of Researchers\\in the Internet Age}
\author[1]{Alberto Bartoli\thanks{Alberto Bartoli: \url{http://bartoli.inginf.units.it}}}
\author[1]{Eric Medvet\thanks{Eric Medvet: \url{http://medvet.inginf.units.it}}}
\affil[1]{DIA, University of Trieste, Italy}
\date{}

\maketitle

\begin{abstract}
Bibliometric measures derived from citation counts are increasingly being used as a research evaluation tool.
Their strengths and weaknesses have been widely analyzed in the literature and are often subject of vigorous debate.
We believe there are a few fundamental issues related to the impact of the web on bibliographic practices that are not taken into account with the importance they deserve.
We focus on evaluation of researchers, but several of our arguments may be applied also to evaluation of research institutions as well as of journals and conferences.
\end{abstract}

\section{Introduction}
Bibliometric measures derived from citation counts are increasingly being used as a research evaluation tool.
Their strengths and weaknesses have been widely analyzed in the literature~\cite{bornmann2008} and are often subject of vigorous debate, also and especially in the computing research community~\cite{meyer2009}.
In this short paper we highlight two fundamental issues that, in our opinion, are not taken into account with the importance they deserve.
\begin{enumerate}
    \item Causal relations between high quality work and citation counts are vanishing.
    \item Uncitedness of high quality work is not an exceptional case.
\end{enumerate}
In the next sections we elaborate on these claims and on their interactions.
We focus on evaluation of researchers, but several of our arguments may be applied also to evaluation of research institutions as well as of journals and conferences.

\section{Causal relations between high quality work and citation counts are vanishing}
The bibliographic practices of researchers have changed radically in recent years, in particular, concerning the discovery of relevant works~\cite{schonfeld2010}.
Prior to the diffusion of the web, researchers were forced to focus their reading and searching efforts on a well-defined set of journals and proceedings.
As a result, papers published in these venues had clearly a much higher chance of being cited than papers published elsewhere.
On the other hand, there was a strong competition for publishing in these venues precisely because of their much higher chance to be read and, thus, cited.
In fact, it is this observation that ultimately justifies the use of citation counts as a surrogate for measuring the quality of publication venues.

Today researchers query search engines for the keyword sets they deem relevant.
Search engines rank the result set and people don't look further than the first pages of hits.
In other words, all papers related to the chosen keyword set have essentially equal chances of being placed to the attention of a researcher, irrespective of the rigor of the reviewing process, of the acceptance criteria enforced by editors and so on.
As a result, these factors no longer play a crucial role in the construction of the set of papers that may potentially be cited.

Academic search engines obviously do attempt to take quality---in a broad sense---into account when constructing and ranking a result set.
On the other hand, an algorithm for reliably quantifying quality of a scholar article has yet to be found.
Moreover, result construction and ranking has to mix quality estimate with content relevance, but the details of this procedure are engine-dependent and unknown.
In practice, quality estimate may easily be obfuscated by content relevance.
Citation counts appear to play a major role in ranking algorithms, but their role with respect to content relevance is, again, engine-dependent and unknown.
Besides, the recent recommendation service by Google Scholar seems to be entirely content based.
In summary, papers floating in the ocean of low citation counts are read with a process that depends on many factors, in which quality plays an increasingly less relevant role. 

Several recent results in bibliometric research confirm that the causal relation between high quality and citation count is vanishing: \emph{``Throughout most of the 20th century, papers' citation rates were increasingly linked to their respective journals' Impact Factors.
However, since 1990, the advent of the digital age, [\dots] the proportion of highly cited papers coming from highly cited journals has been decreasing, and accordingly, the proportion of highly cited papers not coming from highly cited journals has also been increasing''}~\cite{lozano2012}.
This result strongly suggests that reading efforts are increasingly less focused on selected publications.

Another factor which impacts on reading habits---and, eventually, how researchers cite---is the availability of papers.
\cite{gargouri2010}~focused on papers whose authors have supplemented sub\-scription-based access to the publisher's version by self-archiving their own copy and making it freely accessible on the web.
This study analyzed more than 27000 articles published in nearly 2000 journals in years 2000--2006 and demonstrated that papers made accessible in this Open Access form \emph{``are cited significantly more than papers in the same journal and year that have not been made Open Access''}.
This very same phenomenon had been already pointed out back in 2001, by analyzing 120.000 papers published in computer science conferences~\cite{lawrence2001}: \emph{``The results are dramatic, showing a clear correlation between the number of times an article is cited and the probability that the article is (freely) online''}.
Morever \emph{``If we assume that articles published in the same venue (proceedings for a given year) are of similar quality, then the analysis by venue suggests that online articles are more highly cited because of their easier availability''}.
These results confirm that reading efforts are becoming more influenced by ease of web access than by the need of reviewing the literature more accurately.

One might argue that authors are competent enough to select what is worth citing and what is not.
The point is that an author can hardly have the experience or specific competence of a reviewer.
In the past, authors that focused on high quality venues for selecting what to cite benefited automatically from the informed filtering by editorial boards.
Today, this form of quality certification is increasingly overwhelmed by such radically different factors as ease of web discovery, ease of access and content relevance.
The likelihood that a paper takes more citations of another paper which has gone through a much more rigorous review is much higher today than it was used to be 10 or 20 years ago.

\section{Uncitedness of high quality work is not an exceptional case}
Every year, a significant percentage of papers that should be considered as being of ``high quality'' under any metric or human judgment, either never get cited at all or take just a few citations.
In other words, high quality papers are not immune from the well known observation that a few papers take most of the citations~\cite{bornmann2008}.

To illustrate this often overlooked phenomenon, we considered the Computer Science field and collected the citation counts for all papers published in some top-level venues of the respective subfields.
Table~\ref{tab:venues} shows the 8 venues we chose, along with their acronyms.
\begin{table*}
    \centering
    \begin{tabular}{lll}
        \hline
        Publication venue & Type & Acronym \\
        \hline
        ACM Transactions on Computer Systems & journal & TOCS \\
        ACM Transactions on Internet Technology & journal & TOIT \\
        IEEE Transactions on Knowledge and Data Engineering & journal & TKDE \\
        IEEE Transactions on Software Engineering & journal & TSE \\
        ACM Symposium on Operating Systems Principles & conference & SOSP \\
        IEEE Symposium on Security and Privacy & conference & S\&P \\
        Internet Measurement Conference & conference & IMC \\
        Symposium on Principles of Distributed Computing & conference & PODC \\
        \hline
    \end{tabular}
    \caption{\label{tab:venues}Publication venues and corresponding acronyms.}
\end{table*}
Certainly, the list is not exhaustive and not every paper published in these venues is groundbreaking.
It is equally true, though, that every such paper has undergone through a state-of-the-art reviewing process, i.e., a careful analysis by several independent people that are recognized as experts by the relevant scientific community.
Any reasonable assessment of research quality must necessarily assign to those publications a weight that is not negligible.
We focused on years 2000--2009: a range sufficiently small to be analyzed easily and sufficiently large to filter possible anomalies out.
It also extends in the past sufficiently enough to allow each paper to be cited.
We collected the data from Microsoft Academic Search API\footnote{\url{http://academic.research.microsoft.com}} during November, 2012.
Since our aim is not obtaining a ``perfect'' measure but just a high-level assessment, possible weaknesses in terms of potential undercitations may be tolerated.
Table~\ref{tab:cit-all} shows, for each year and publication venues, the percentage of papers which have taken 1 or 0 citations---we consider \emph{all} the citations received up to the data collection date, i.e., up to 2012.
For example, half of the papers at IMC 2004 have been cited once or never at all.
\begin{table*}
    \centering
    \begin{tabular}{lrrrrrrrrrr}
        \hline
        & 2000 & 2001 & 2002 & 2003 & 2004 & 2005 & 2006 & 2007 & 2008 & 2009 \\ 
        \hline
        TOCS & 0 & 16 & 7 & 0 & 17 & 0 & 0 & 8 & 20 & 12 \\ 
        TOIT & 0 & 14 & 14 & 0 & 12 & 5 & 11 & 17 & 19 & 25 \\ 
        TKDE & 14 & 18 & 10 & 9 & 18 & 14 & 19 & 26 & 25 & 39 \\ 
        TSE & 14 & 13 & 10 & 16 & 29 & 17 & 25 & 25 & 28 & 21 \\ 
        SOSP &  & 0 &  & 6 &  & 57 &  & 4 &  & 12 \\ 
        S\&P & 10 & 4 & 0 & 8 & 14 & 5 & 3 & 9 & 3 & 7 \\ 
        IMC & 0 &  & 33 & 33 & 50 & 8 & 3 & 5 & 9 & 17 \\ 
        PODC & 35 & 17 & 18 & 19 & 33 & 21 & 11 & 32 & 44 & 54 \\ 
        \hline
    \end{tabular}
    \caption{
        \label{tab:cit-all}
        Percentage of papers which received, up to November, 2012, 0 or 1 citations, per publication venue.
        Blank table cells correspond to unavailable data (e.g., a conference was not held during a given year).
    }
\end{table*}

Table~\ref{tab:cit-2year} shows, for each year and publication venues, the percentage of papers which have taken 0 citations within 2 year upon publication.
For example, 3 on 4 papers published at PODC 2004 received 0 citations from other papers published in years 2004, 2005 and 2006.
\begin{table*}
    \centering
    \begin{tabular}{rrrrrrrrrrr}
        \hline
        & 2000 & 2001 & 2002 & 2003 & 2004 & 2005 & 2006 & 2007 & 2008 & 2009 \\ 
        \hline
        TOCS & 11 & 26 & 13 & 15 & 17 & 0 & 0 & 8 & 20 & 12 \\ 
        TOIT & 100 & 71 & 71 & 31 & 41 & 41 & 63 & 59 & 81 & 56 \\ 
        TKDE & 80 & 78 & 75 & 67 & 67 & 69 & 76 & 76 & 73 & 72 \\ 
        TSE & 72 & 57 & 64 & 65 & 72 & 64 & 72 & 68 & 70 & 64 \\ 
        SOSP &  & 56 &  & 44 &  & 96 &  & 62 &  & 50 \\ 
        S\&P & 59 & 40 & 19 & 31 & 27 & 19 & 32 & 38 & 24 & 56 \\ 
        IMC & 100 &  & 100 & 83 & 100 & 68 & 61 & 83 & 72 & 90 \\ 
        PODC & 91 & 72 & 59 & 55 & 75 & 70 & 49 & 71 & 80 & 85 \\ 
        \hline
    \end{tabular}
    \caption{
        \label{tab:cit-2year}
        Percentage of papers which received 0 citations within 2 years upon publication, per publication venue.
        Blank table cells correspond to unavailable data (e.g., a conference was not held during a given year).
    }
\end{table*}

\begin{figure}
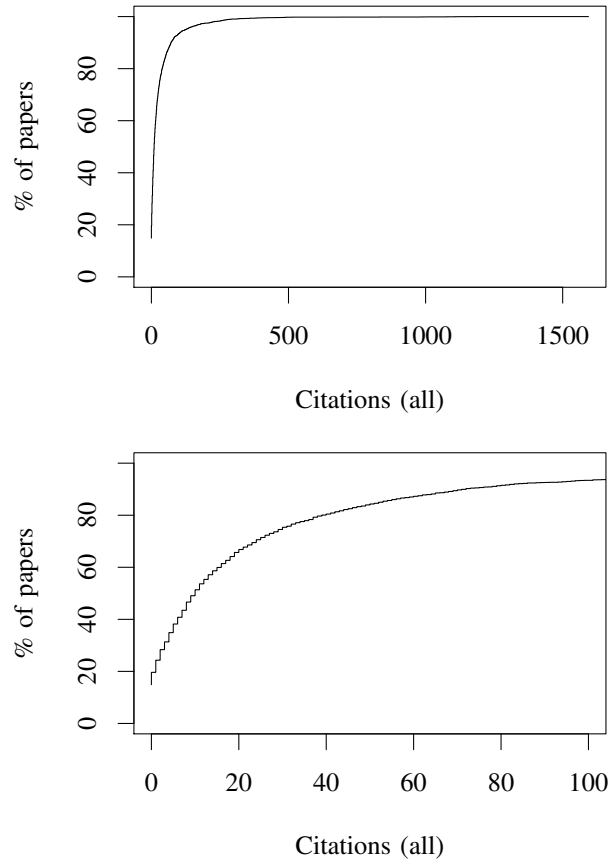

    \centering
    \begin{subfigure}{1\columnwidth}
        \centering
        \vspace{-1.1cm}
        \include{figures/all.all}
        \vspace{-1cm}
    \end{subfigure}
    \begin{subfigure}{1\columnwidth}
        \centering
        \vspace{-1.1cm}
        \include{figures/all.100}
        \vspace{-0.9cm}
    \end{subfigure}
    \caption{
        \label{fig:cit-ecdf-all}
        Percentage of papers which received less than $x$ citations, counting all citations received up to November, 2012.
        The chart below shows the detail for $x \le 100$.
    }
\end{figure}

\begin{figure}
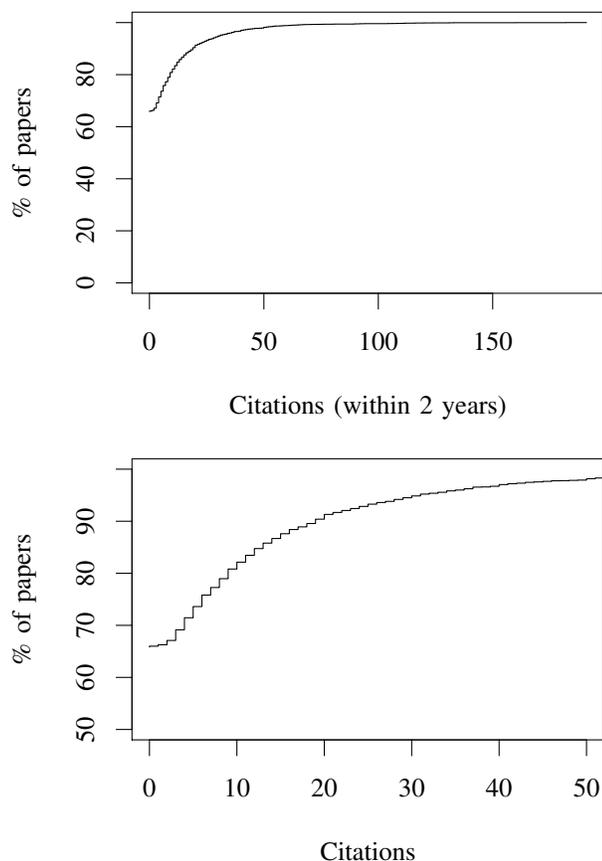

    \centering
    \begin{subfigure}{1\columnwidth}
        \centering
        \vspace{-1.1cm}
        \include{figures/0-2.all}
        \vspace{-1cm}
    \end{subfigure}
    \begin{subfigure}{1\columnwidth}
        \centering
        \vspace{-1.1cm}
        \include{figures/0-2.50}
        \vspace{-0.9cm}
    \end{subfigure}
    \caption{
        \label{fig:cit-ecdf-0-2}
        Percentage of papers which received less than $x$ citations, counting all citations received within 2 years upon publication.
        The chart below shows the detail for $x \le 50$.
    }
\end{figure}

The same data is plotted in Figures~\ref{fig:cit-ecdf-all} and~\ref{fig:cit-ecdf-0-2} which show the percentage of papers which received less than a given number of citations, counting all citations and citations received within 2 years upon publication, respectively.
For example, Figure~\ref{fig:cit-ecdf-0-2} shows that more than 60\% of papers is never cited within 2 years upon publication.
Other elaborations of this data are available at \url{http://machinelearning.inginf.units.it/data-and-tools/paper-citations-for-important-cs-venues}.

The implication of these distributions is clear: when evaluation of researchers is based on citation counts, a large percentage of papers that are definitely high quality have a remarkable chance of contributing very little, if at all, to the assessment.
The Italian Government, for example, has recently established necessary bibliometric-based conditions that faculty members have to satisfy in order to be eligible for a promotion.
The data are extracted from Scopus and Web of Science (it has been argued that these sources are not adequate for computer science~\cite{meyer2009}).
A candidate has to satisfy at least two of these three criteria: number of journal publications, citation counts, contemporary h-index (a variant of the h-index in which the number of citations is weighed by the number of years elapsed from the time of the publication)---we omit the details for brevity.
Table~\ref{tab:cit-all} suggests that a significant percentage of TSE papers, for example, may contribute very little to the second and third criterion, if at all.
Those papers might even contribute less than papers at any of the many Scopus-indexed conferences whose quality is well below TSE under any informed human judgment.
Concerning the first criterion an SOSP paper, for example, does not count at all.

\section{Discussion}
An ecosystem where incentives are heavily based on citation counts and where discovery of relevant literature is mostly based on content relevance and citation counts, could evolve along unexpected paths and challenge several of the assumptions traditionally taken for granted.

Submitting to a top-level journal may lead to several review rounds, each requiring a significant amount of work, with a turnaround time that may take years and conclude with a rejection.
From the point of view of authors that have to accumulate citation counts, submitting to such a journal may be an irrational move: high cost and high risk with little, if any, advantage in return.
There is definitely a strong advantage in submitting at journals with easier-to-satisfy editorial boards.
As soon as a paper has been published on some ``indexed'' venue, it becomes ready to appear in search engine results and, thus, ready to be cited.

Based on this observation, editorial boards and commercial publishers could have an incentive in lowering the bar while maintaining a sort of ``decent'' quality.
In recent years we have observed several cases of journals---by commercial publishers---whose number of published articles per year has grown very quickly by more than one order of magnitude.
We have also observed a similar trend regarding several collections of computer science conference.
We prefer not to point at any specific case because we do not have any authority to judge the quality of these publications.
It is our opinion, though, that the citation game is playing an important role here and we merely invite the community to reflect on this trend.
Incentives in avoiding high quality reviewing processes may have disrupting effects in the long term that can be imagined easily.

The writing process itself could change radically:
\begin{inparaenum}[(i)]
    \item papers should be written according to academic search engine optimization techniques~\cite{elsevier2012,beel2010}; and
    \item title and abstract should be chosen after a careful analysis of candidate keywords in Google Trends or similar tools.
\end{inparaenum}
Diversity of research topics could greatly suffer, as exploratory research in areas that are not mainstream would be penalized at evaluation time.
Finally, spam and security problems could become a serious issue also in research platforms, as there would be an enormous and widespread interest in manipulating publicly available pdf files for artificially increasing citation counts~\cite{beel2010google,cyril2010}.
Even a transitory increase would suffice, as it would allow the fraudulently promoted paper to emerge from the ocean of low citation counts and start collecting legitimate citations.

We hope the issues outlined in this viewpoint will not be taken lightly.
Recognizing outstanding researchers is relatively easy.
Ensuring that the myriad of serious researchers that happen not to be outstanding be given the chances they deserve, while being assessed fairly is much more difficult.
In our modest opinion, bibliometric-based measures fail exactly in this second category and their increased importance and diffusion incur significant risks for the whole community.

\bibliographystyle{plain}
\bibliography{main}

\end{document}

%% file: figures/all.all.tex
\begin{tikzpicture}[x=1pt,y=1pt]
\definecolor[named]{fillColor}{rgb}{1.00,1.00,1.00}
\path[use as bounding box,fill=fillColor,fill opacity=0.00] (0,0) rectangle (252.94,216.81);
\begin{scope}
\path[clip] ( 49.20, 61.20) rectangle (227.75,167.61);
\definecolor[named]{drawColor}{rgb}{0.00,0.00,0.00}

\path[draw=drawColor,line width= 0.4pt,line join=round,line cap=round] ( 55.81, 79.80) --
    ( 55.81, 84.49) --
	( 55.92, 84.49) --
	( 55.92, 89.08) --
	( 56.02, 89.08) --
	( 56.02, 93.08) --
	( 56.12, 93.08) --
	( 56.12, 96.02) --
	( 56.23, 96.02) --
	( 56.23, 99.52) --
	( 56.33, 99.52) --
	( 56.33,102.78) --
	( 56.44,102.78) --
	( 56.44,105.33) --
	( 56.54,105.33) --
	( 56.54,107.98) --
	( 56.64,107.98) --
	( 56.64,111.10) --
	( 56.75,111.10) --
	( 56.75,113.49) --
	( 56.85,113.49) --
	( 56.85,115.74) --
	( 56.95,115.74) --
	( 56.95,118.00) --
	( 57.06,118.00) --
	( 57.06,119.67) --
	( 57.16,119.67) --
	( 57.16,121.44) --
	( 57.26,121.44) --
	( 57.26,122.93) --
	( 57.37,122.93) --
	( 57.37,124.20) --
	( 57.47,124.20) --
	( 57.47,125.66) --
	( 57.58,125.66) --
	( 57.58,126.88) --
	( 57.68,126.88) --
	( 57.68,128.33) --
	( 57.78,128.33) --
	( 57.78,129.87) --
	( 57.89,129.87) --
	( 57.89,130.99) --
	( 57.99,130.99) --
	( 57.99,131.89) --
	( 58.09,131.89) --
	( 58.09,132.71) --
	( 58.20,132.71) --
	( 58.20,133.61) --
	( 58.30,133.61) --
	( 58.30,134.70) --
	( 58.41,134.70) --
	( 58.41,135.54) --
	( 58.51,135.54) --
	( 58.51,136.39) --
	( 58.61,136.39) --
	( 58.61,137.08) --
	( 58.72,137.08) --
	( 58.72,137.69) --
	( 58.82,137.69) --
	( 58.82,138.57) --
	( 58.92,138.57) --
	( 58.92,139.41) --
	( 59.03,139.41) --
	( 59.03,139.81) --
	( 59.13,139.81) --
	( 59.13,140.61) --
	( 59.24,140.61) --
	( 59.24,141.14) --
	( 59.34,141.14) --
	( 59.34,141.56) --
	( 59.44,141.56) --
	( 59.44,141.91) --
	( 59.55,141.91) --
	( 59.55,142.33) --
	( 59.65,142.33) --
	( 59.65,143.21) --
	( 59.75,143.21) --
	( 59.75,143.66) --
	( 59.86,143.66) --
	( 59.86,144.03) --
	( 59.96,144.03) --
	( 59.96,144.40) --
	( 60.07,144.40) --
	( 60.07,144.80) --
	( 60.17,144.80) --
	( 60.17,145.27) --
	( 60.27,145.27) --
	( 60.27,145.64) --
	( 60.38,145.64) --
	( 60.38,146.12) --
	( 60.48,146.12) --
	( 60.48,146.44) --
	( 60.58,146.44) --
	( 60.58,146.84) --
	( 60.69,146.84) --
	( 60.69,147.21) --
	( 60.79,147.21) --
	( 60.79,147.45) --
	( 60.89,147.45) --
	( 60.89,147.87) --
	( 61.00,147.87) --
	( 61.00,148.22) --
	( 61.10,148.22) --
	( 61.10,148.48) --
	( 61.21,148.48) --
	( 61.21,148.88) --
	( 61.31,148.88) --
	( 61.31,149.28) --
	( 61.41,149.28) --
	( 61.41,149.54) --
	( 61.52,149.54) --
	( 61.52,149.89) --
	( 61.62,149.89) --
	( 61.62,150.18) --
	( 61.72,150.18) --
	( 61.72,150.47) --
	( 61.83,150.47) --
	( 61.83,150.63) --
	( 61.93,150.63) --
	( 61.93,150.87) --
	( 62.04,150.87) --
	( 62.04,151.13) --
	( 62.14,151.13) --
	( 62.14,151.34) --
	( 62.24,151.34) --
	( 62.24,151.66) --
	( 62.35,151.66) --
	( 62.35,151.77) --
	( 62.45,151.77) --
	( 62.45,152.01) --
	( 62.55,152.01) --
	( 62.55,152.40) --
	( 62.66,152.40) --
	( 62.66,152.48) --
	( 62.76,152.48) --
	( 62.76,152.72) --
	( 62.87,152.72) --
	( 62.87,152.96) --
	( 62.97,152.96) --
	( 62.97,153.25) --
	( 63.07,153.25) --
	( 63.07,153.54) --
	( 63.18,153.54) --
	( 63.18,153.76) --
	( 63.28,153.76) --
	( 63.28,154.05) --
	( 63.38,154.05) --
	( 63.38,154.23) --
	( 63.49,154.23) --
	( 63.49,154.31) --
	( 63.59,154.31) --
	( 63.59,154.44) --
	( 63.70,154.44) --
	( 63.70,154.63) --
	( 63.80,154.63) --
	( 63.80,154.68) --
	( 63.90,154.68) --
	( 63.90,154.89) --
	( 64.01,154.89) --
	( 64.01,155.11) --
	( 64.11,155.11) --
	( 64.11,155.29) --
	( 64.21,155.29) --
	( 64.21,155.43) --
	( 64.32,155.43) --
	( 64.32,155.64) --
	( 64.42,155.64) --
	( 64.42,155.82) --
	( 64.52,155.82) --
	( 64.52,155.93) --
	( 64.63,155.93) --
	( 64.63,156.01) --
	( 64.73,156.01) --
	( 64.73,156.17) --
	( 64.94,156.17) --
	( 64.94,156.27) --
	( 65.04,156.27) --
	( 65.04,156.33) --
	( 65.15,156.33) --
	( 65.15,156.38) --
	( 65.25,156.38) --
	( 65.25,156.43) --
	( 65.46,156.43) --
	( 65.46,156.51) --
	( 65.56,156.51) --
	( 65.56,156.64) --
	( 65.67,156.64) --
	( 65.67,156.75) --
	( 65.77,156.75) --
	( 65.77,156.88) --
	( 65.87,156.88) --
	( 65.87,157.04) --
	( 65.98,157.04) --
	( 65.98,157.12) --
	( 66.08,157.12) --
	( 66.08,157.17) --
	( 66.29,157.17) --
	( 66.29,157.36) --
	( 66.39,157.36) --
	( 66.39,157.39) --
	( 66.50,157.39) --
	( 66.50,157.44) --
	( 66.60,157.44) --
	( 66.60,157.63) --
	( 66.70,157.63) --
	( 66.70,157.68) --
	( 66.81,157.68) --
	( 66.81,157.76) --
	( 66.91,157.76) --
	( 66.91,157.84) --
	( 67.01,157.84) --
	( 67.01,157.94) --
	( 67.12,157.94) --
	( 67.12,157.97) --
	( 67.22,157.97) --
	( 67.22,158.05) --
	( 67.32,158.05) --
	( 67.32,158.18) --
	( 67.43,158.18) --
	( 67.43,158.23) --
	( 67.53,158.23) --
	( 67.53,158.31) --
	( 67.64,158.31) --
	( 67.64,158.34) --
	( 67.74,158.34) --
	( 67.74,158.39) --
	( 68.05,158.39) --
	( 68.05,158.42) --
	( 68.15,158.42) --
	( 68.15,158.45) --
	( 68.26,158.45) --
	( 68.26,158.53) --
	( 68.36,158.53) --
	( 68.36,158.63) --
	( 68.57,158.63) --
	( 68.57,158.69) --
	( 68.67,158.69) --
	( 68.67,158.74) --
	( 68.78,158.74) --
	( 68.78,158.77) --
	( 68.88,158.77) --
	( 68.88,158.84) --
	( 68.98,158.84) --
	( 68.98,158.90) --
	( 69.09,158.90) --
	( 69.09,158.95) --
	( 69.19,158.95) --
	( 69.19,158.98) --
	( 69.30,158.98) --
	( 69.30,159.00) --
	( 69.40,159.00) --
	( 69.40,159.08) --
	( 69.50,159.08) --
	( 69.50,159.16) --
	( 69.61,159.16) --
	( 69.61,159.19) --
	( 69.71,159.19) --
	( 69.71,159.27) --
	( 69.81,159.27) --
	( 69.81,159.32) --
	( 69.92,159.32) --
	( 69.92,159.37) --
	( 70.02,159.37) --
	( 70.02,159.40) --
	( 70.23,159.40) --
	( 70.23,159.51) --
	( 70.44,159.51) --
	( 70.44,159.59) --
	( 70.54,159.59) --
	( 70.54,159.67) --
	( 70.64,159.67) --
	( 70.64,159.69) --
	( 70.85,159.69) --
	( 70.85,159.72) --
	( 71.06,159.72) --
	( 71.06,159.77) --
	( 71.16,159.77) --
	( 71.16,159.83) --
	( 71.27,159.83) --
	( 71.27,159.88) --
	( 71.37,159.88) --
	( 71.37,159.98) --
	( 71.58,159.98) --
	( 71.58,160.01) --
	( 71.68,160.01) --
	( 71.68,160.06) --
	( 71.99,160.06) --
	( 71.99,160.09) --
	( 72.10,160.09) --
	( 72.10,160.12) --
	( 72.20,160.12) --
	( 72.20,160.17) --
	( 72.41,160.17) --
	( 72.41,160.20) --
	( 72.61,160.20) --
	( 72.61,160.22) --
	( 72.72,160.22) --
	( 72.72,160.28) --
	( 72.82,160.28) --
	( 72.82,160.38) --
	( 72.93,160.38) --
	( 72.93,160.41) --
	( 73.03,160.41) --
	( 73.03,160.49) --
	( 73.24,160.49) --
	( 73.24,160.54) --
	( 73.34,160.54) --
	( 73.34,160.57) --
	( 73.44,160.57) --
	( 73.44,160.62) --
	( 73.65,160.62) --
	( 73.65,160.67) --
	( 74.07,160.67) --
	( 74.07,160.70) --
	( 74.17,160.70) --
	( 74.17,160.78) --
	( 74.38,160.78) --
	( 74.38,160.89) --
	( 74.48,160.89) --
	( 74.48,160.91) --
	( 74.58,160.91) --
	( 74.58,160.97) --
	( 74.69,160.97) --
	( 74.69,160.99) --
	( 75.10,160.99) --
	( 75.10,161.02) --
	( 75.41,161.02) --
	( 75.41,161.04) --
	( 75.52,161.04) --
	( 75.52,161.07) --
	( 75.73,161.07) --
	( 75.73,161.10) --
	( 75.83,161.10) --
	( 75.83,161.15) --
	( 76.04,161.15) --
	( 76.04,161.18) --
	( 76.56,161.18) --
	( 76.56,161.20) --
	( 77.07,161.20) --
	( 77.07,161.23) --
	( 77.39,161.23) --
	( 77.39,161.26) --
	( 77.49,161.26) --
	( 77.49,161.28) --
	( 77.80,161.28) --
	( 77.80,161.31) --
	( 77.90,161.31) --
	( 77.90,161.36) --
	( 78.01,161.36) --
	( 78.01,161.39) --
	( 78.11,161.39) --
	( 78.11,161.44) --
	( 78.21,161.44) --
	( 78.21,161.47) --
	( 78.53,161.47) --
	( 78.53,161.50) --
	( 78.63,161.50) --
	( 78.63,161.52) --
	( 78.73,161.52) --
	( 78.73,161.55) --
	( 78.84,161.55) --
	( 78.84,161.63) --
	( 79.04,161.63) --
	( 79.04,161.68) --
	( 79.15,161.68) --
	( 79.15,161.73) --
	( 79.67,161.73) --
	( 79.67,161.76) --
	( 79.77,161.76) --
	( 79.77,161.81) --
	( 80.08,161.81) --
	( 80.08,161.84) --
	( 80.60,161.84) --
	( 80.60,161.87) --
	( 80.91,161.87) --
	( 80.91,161.89) --
	( 81.02,161.89) --
	( 81.02,161.95) --
	( 81.74,161.95) --
	( 81.74,162.00) --
	( 81.95,162.00) --
	( 81.95,162.08) --
	( 82.05,162.08) --
	( 82.05,162.10) --
	( 82.16,162.10) --
	( 82.16,162.13) --
	( 82.57,162.13) --
	( 82.57,162.18) --
	( 82.67,162.18) --
	( 82.67,162.21) --
	( 82.78,162.21) --
	( 82.78,162.26) --
	( 82.99,162.26) --
	( 82.99,162.29) --
	( 83.30,162.29) --
	( 83.30,162.32) --
	( 83.40,162.32) --
	( 83.40,162.34) --
	( 83.61,162.34) --
	( 83.61,162.40) --
	( 83.82,162.40) --
	( 83.82,162.42) --
	( 83.92,162.42) --
	( 83.92,162.45) --
	( 84.13,162.45) --
	( 84.13,162.48) --
	( 84.44,162.48) --
	( 84.44,162.50) --
	( 84.65,162.50) --
	( 84.65,162.56) --
	( 84.75,162.56) --
	( 84.75,162.58) --
	( 85.27,162.58) --
	( 85.27,162.64) --
	( 85.68,162.64) --
	( 85.68,162.66) --
	( 85.89,162.66) --
	( 85.89,162.69) --
	( 85.99,162.69) --
	( 85.99,162.71) --
	( 86.41,162.71) --
	( 86.41,162.74) --
	( 88.17,162.74) --
	( 88.17,162.77) --
	( 88.28,162.77) --
	( 88.28,162.79) --
	( 88.48,162.79) --
	( 88.48,162.82) --
	( 88.69,162.82) --
	( 88.69,162.85) --
	( 88.79,162.85) --
	( 88.79,162.87) --
	( 89.21,162.87) --
	( 89.21,162.90) --
	( 89.93,162.90) --
	( 89.93,162.93) --
	( 91.18,162.93) --
	( 91.18,162.95) --
	( 91.70,162.95) --
	( 91.70,162.98) --
	( 92.84,162.98) --
	( 92.84,163.01) --
	( 92.94,163.01) --
	( 92.94,163.03) --
	( 93.15,163.03) --
	( 93.15,163.06) --
	( 94.91,163.06) --
	( 94.91,163.09) --
	( 95.33,163.09) --
	( 95.33,163.11) --
	( 95.64,163.11) --
	( 95.64,163.14) --
	( 96.05,163.14) --
	( 96.05,163.17) --
	( 97.40,163.17) --
	( 97.40,163.19) --
	( 97.92,163.19) --
	( 97.92,163.22) --
	( 98.85,163.22) --
	( 98.85,163.24) --
	( 99.48,163.24) --
	( 99.48,163.27) --
	(102.69,163.27) --
	(102.69,163.30) --
	(104.45,163.30) --
	(104.45,163.32) --
	(105.28,163.32) --
	(105.28,163.35) --
	(106.22,163.35) --
	(106.22,163.38) --
	(106.63,163.38) --
	(106.63,163.40) --
	(106.84,163.40) --
	(106.84,163.43) --
	(109.23,163.43) --
	(109.23,163.46) --
	(126.75,163.46) --
	(126.75,163.48) --
	(142.10,163.48) --
	(142.10,163.51) --
	(157.87,163.51) --
	(157.87,163.54) --
	(160.04,163.54) --
	(160.04,163.56) --
	(167.41,163.56) --
	(167.41,163.59) --
	(170.73,163.59) --
	(170.73,163.62) --
	(171.87,163.62) --
	(171.87,163.64) --
	(178.82,163.64) --
	(178.82,163.67) --
	(221.13,163.67);
\end{scope}
\begin{scope}
\path[clip] (  0.00,  0.00) rectangle (252.94,216.81);
\definecolor[named]{drawColor}{rgb}{0.00,0.00,0.00}

\path[draw=drawColor,line width= 0.4pt,line join=round,line cap=round] ( 55.81, 61.20) -- (211.38, 61.20);

\path[draw=drawColor,line width= 0.4pt,line join=round,line cap=round] ( 55.81, 61.20) -- ( 55.81, 55.20);

\path[draw=drawColor,line width= 0.4pt,line join=round,line cap=round] (107.67, 61.20) -- (107.67, 55.20);

\path[draw=drawColor,line width= 0.4pt,line join=round,line cap=round] (159.53, 61.20) -- (159.53, 55.20);

\path[draw=drawColor,line width= 0.4pt,line join=round,line cap=round] (211.38, 61.20) -- (211.38, 55.20);

\node[text=drawColor,anchor=base,inner sep=0pt, outer sep=0pt, scale=  1.00] at ( 55.81, 39.60) {0};

\node[text=drawColor,anchor=base,inner sep=0pt, outer sep=0pt, scale=  1.00] at (107.67, 39.60) {500};

\node[text=drawColor,anchor=base,inner sep=0pt, outer sep=0pt, scale=  1.00] at (159.53, 39.60) {1000};

\node[text=drawColor,anchor=base,inner sep=0pt, outer sep=0pt, scale=  1.00] at (211.38, 39.60) {1500};

\path[draw=drawColor,line width= 0.4pt,line join=round,line cap=round] ( 49.20, 65.14) -- ( 49.20,163.67);

\path[draw=drawColor,line width= 0.4pt,line join=round,line cap=round] ( 49.20, 65.14) -- ( 43.20, 65.14);

\path[draw=drawColor,line width= 0.4pt,line join=round,line cap=round] ( 49.20, 84.85) -- ( 43.20, 84.85);

\path[draw=drawColor,line width= 0.4pt,line join=round,line cap=round] ( 49.20,104.55) -- ( 43.20,104.55);

\path[draw=drawColor,line width= 0.4pt,line join=round,line cap=round] ( 49.20,124.26) -- ( 43.20,124.26);

\path[draw=drawColor,line width= 0.4pt,line join=round,line cap=round] ( 49.20,143.96) -- ( 43.20,143.96);

\path[draw=drawColor,line width= 0.4pt,line join=round,line cap=round] ( 49.20,163.67) -- ( 43.20,163.67);

\node[text=drawColor,rotate= 90.00,anchor=base,inner sep=0pt, outer sep=0pt, scale=  1.00] at ( 34.80, 65.14) {0};

\node[text=drawColor,rotate= 90.00,anchor=base,inner sep=0pt, outer sep=0pt, scale=  1.00] at ( 34.80, 84.85) {20};

\node[text=drawColor,rotate= 90.00,anchor=base,inner sep=0pt, outer sep=0pt, scale=  1.00] at ( 34.80,104.55) {40};

\node[text=drawColor,rotate= 90.00,anchor=base,inner sep=0pt, outer sep=0pt, scale=  1.00] at ( 34.80,124.26) {60};

\node[text=drawColor,rotate= 90.00,anchor=base,inner sep=0pt, outer sep=0pt, scale=  1.00] at ( 34.80,143.96) {80};

\path[draw=drawColor,line width= 0.4pt,line join=round,line cap=round] ( 49.20, 61.20) --
	(227.75, 61.20) --
	(227.75,167.61) --
	( 49.20,167.61) --
	( 49.20, 61.20);
\end{scope}
\begin{scope}
\path[clip] (  0.00,  0.00) rectangle (252.94,216.81);
\definecolor[named]{drawColor}{rgb}{0.00,0.00,0.00}

\node[text=drawColor,anchor=base,inner sep=0pt, outer sep=0pt, scale=  1.00] at (138.47, 15.60) {Citations (all)};

\node[text=drawColor,rotate= 90.00,anchor=base,inner sep=0pt, outer sep=0pt, scale=  1.00] at ( 10.80,114.41) {\% of papers};
\end{scope}
\end{tikzpicture}

%% file: figures/all.100.tex
\begin{tikzpicture}[x=1pt,y=1pt]
\definecolor[named]{fillColor}{rgb}{1.00,1.00,1.00}
\path[use as bounding box,fill=fillColor,fill opacity=0.00] (0,0) rectangle (252.94,216.81);
\begin{scope}
\path[clip] ( 49.20, 61.20) rectangle (227.75,167.61);
\definecolor[named]{drawColor}{rgb}{0.00,0.00,0.00}

\path[draw=drawColor,line width= 0.4pt,line join=round,line cap=round] ( 55.81, 79.80) --
	( 55.81, 84.49) --
	( 57.47, 84.49) --
	( 57.47, 89.08) --
	( 59.12, 89.08) --
	( 59.12, 93.08) --
	( 60.77, 93.08) --
	( 60.77, 96.02) --
	( 62.43, 96.02) --
	( 62.43, 99.52) --
	( 64.08, 99.52) --
	( 64.08,102.78) --
	( 65.73,102.78) --
	( 65.73,105.33) --
	( 67.39,105.33) --
	( 67.39,107.98) --
	( 69.04,107.98) --
	( 69.04,111.10) --
	( 70.69,111.10) --
	( 70.69,113.49) --
	( 72.34,113.49) --
	( 72.34,115.74) --
	( 74.00,115.74) --
	( 74.00,118.00) --
	( 75.65,118.00) --
	( 75.65,119.67) --
	( 77.30,119.67) --
	( 77.30,121.44) --
	( 78.96,121.44) --
	( 78.96,122.93) --
	( 80.61,122.93) --
	( 80.61,124.20) --
	( 82.26,124.20) --
	( 82.26,125.66) --
	( 83.92,125.66) --
	( 83.92,126.88) --
	( 85.57,126.88) --
	( 85.57,128.33) --
	( 87.22,128.33) --
	( 87.22,129.87) --
	( 88.88,129.87) --
	( 88.88,130.99) --
	( 90.53,130.99) --
	( 90.53,131.89) --
	( 92.18,131.89) --
	( 92.18,132.71) --
	( 93.84,132.71) --
	( 93.84,133.61) --
	( 95.49,133.61) --
	( 95.49,134.70) --
	( 97.14,134.70) --
	( 97.14,135.54) --
	( 98.80,135.54) --
	( 98.80,136.39) --
	(100.45,136.39) --
	(100.45,137.08) --
	(102.10,137.08) --
	(102.10,137.69) --
	(103.76,137.69) --
	(103.76,138.57) --
	(105.41,138.57) --
	(105.41,139.41) --
	(107.06,139.41) --
	(107.06,139.81) --
	(108.72,139.81) --
	(108.72,140.61) --
	(110.37,140.61) --
	(110.37,141.14) --
	(112.02,141.14) --
	(112.02,141.56) --
	(113.67,141.56) --
	(113.67,141.91) --
	(115.33,141.91) --
	(115.33,142.33) --
	(116.98,142.33) --
	(116.98,143.21) --
	(118.63,143.21) --
	(118.63,143.66) --
	(120.29,143.66) --
	(120.29,144.03) --
	(121.94,144.03) --
	(121.94,144.40) --
	(123.59,144.40) --
	(123.59,144.80) --
	(125.25,144.80) --
	(125.25,145.27) --
	(126.90,145.27) --
	(126.90,145.64) --
	(128.55,145.64) --
	(128.55,146.12) --
	(130.21,146.12) --
	(130.21,146.44) --
	(131.86,146.44) --
	(131.86,146.84) --
	(133.51,146.84) --
	(133.51,147.21) --
	(135.17,147.21) --
	(135.17,147.45) --
	(136.82,147.45) --
	(136.82,147.87) --
	(138.47,147.87) --
	(138.47,148.22) --
	(140.13,148.22) --
	(140.13,148.48) --
	(141.78,148.48) --
	(141.78,148.88) --
	(143.43,148.88) --
	(143.43,149.28) --
	(145.09,149.28) --
	(145.09,149.54) --
	(146.74,149.54) --
	(146.74,149.89) --
	(148.39,149.89) --
	(148.39,150.18) --
	(150.04,150.18) --
	(150.04,150.47) --
	(151.70,150.47) --
	(151.70,150.63) --
	(153.35,150.63) --
	(153.35,150.87) --
	(155.00,150.87) --
	(155.00,151.13) --
	(156.66,151.13) --
	(156.66,151.34) --
	(158.31,151.34) --
	(158.31,151.66) --
	(159.96,151.66) --
	(159.96,151.77) --
	(161.62,151.77) --
	(161.62,152.01) --
	(163.27,152.01) --
	(163.27,152.40) --
	(164.92,152.40) --
	(164.92,152.48) --
	(166.58,152.48) --
	(166.58,152.72) --
	(168.23,152.72) --
	(168.23,152.96) --
	(169.88,152.96) --
	(169.88,153.25) --
	(171.54,153.25) --
	(171.54,153.54) --
	(173.19,153.54) --
	(173.19,153.76) --
	(174.84,153.76) --
	(174.84,154.05) --
	(176.50,154.05) --
	(176.50,154.23) --
	(178.15,154.23) --
	(178.15,154.31) --
	(179.80,154.31) --
	(179.80,154.44) --
	(181.46,154.44) --
	(181.46,154.63) --
	(183.11,154.63) --
	(183.11,154.68) --
	(184.76,154.68) --
	(184.76,154.89) --
	(186.42,154.89) --
	(186.42,155.11) --
	(188.07,155.11) --
	(188.07,155.29) --
	(189.72,155.29) --
	(189.72,155.43) --
	(191.37,155.43) --
	(191.37,155.64) --
	(193.03,155.64) --
	(193.03,155.82) --
	(194.68,155.82) --
	(194.68,155.93) --
	(196.33,155.93) --
	(196.33,156.01) --
	(197.99,156.01) --
	(197.99,156.17) --
	(201.29,156.17) --
	(201.29,156.27) --
	(202.95,156.27) --
	(202.95,156.33) --
	(204.60,156.33) --
	(204.60,156.38) --
	(206.25,156.38) --
	(206.25,156.43) --
	(209.56,156.43) --
	(209.56,156.51) --
	(211.21,156.51) --
	(211.21,156.64) --
	(212.87,156.64) --
	(212.87,156.75) --
	(214.52,156.75) --
	(214.52,156.88) --
	(216.17,156.88) --
	(216.17,157.04) --
	(217.83,157.04) --
	(217.83,157.12) --
	(219.48,157.12) --
	(219.48,157.17) --
	(222.79,157.17) --
	(222.79,157.36) --
	(224.44,157.36) --
	(224.44,157.39) --
	(226.09,157.39) --
	(226.09,157.44) --
	(227.75,157.44) --
	(227.75,157.63) --
	(229.40,157.63) --
	(229.40,157.68) --
	(231.05,157.68) --
	(231.05,157.76) --
	(232.70,157.76) --
	(232.70,157.84) --
	(234.36,157.84) --
	(234.36,157.94) --
	(236.01,157.94) --
	(236.01,157.97) --
	(237.66,157.97) --
	(237.66,158.05) --
	(239.32,158.05) --
	(239.32,158.18) --
	(240.97,158.18) --
	(240.97,158.23) --
	(242.62,158.23) --
	(242.62,158.31) --
	(244.28,158.31) --
	(244.28,158.34) --
	(245.93,158.34) --
	(245.93,158.39) --
	(250.89,158.39) --
	(250.89,158.42) --
	(252.54,158.42) --
	(252.54,158.45) --
	(252.94,158.45);
\end{scope}
\begin{scope}
\path[clip] (  0.00,  0.00) rectangle (252.94,216.81);
\definecolor[named]{drawColor}{rgb}{0.00,0.00,0.00}

\path[draw=drawColor,line width= 0.4pt,line join=round,line cap=round] ( 55.81, 61.20) -- (221.13, 61.20);

\path[draw=drawColor,line width= 0.4pt,line join=round,line cap=round] ( 55.81, 61.20) -- ( 55.81, 55.20);

\path[draw=drawColor,line width= 0.4pt,line join=round,line cap=round] ( 88.88, 61.20) -- ( 88.88, 55.20);

\path[draw=drawColor,line width= 0.4pt,line join=round,line cap=round] (121.94, 61.20) -- (121.94, 55.20);

\path[draw=drawColor,line width= 0.4pt,line join=round,line cap=round] (155.00, 61.20) -- (155.00, 55.20);

\path[draw=drawColor,line width= 0.4pt,line join=round,line cap=round] (188.07, 61.20) -- (188.07, 55.20);

\path[draw=drawColor,line width= 0.4pt,line join=round,line cap=round] (221.13, 61.20) -- (221.13, 55.20);

\node[text=drawColor,anchor=base,inner sep=0pt, outer sep=0pt, scale=  1.00] at ( 55.81, 39.60) {0};

\node[text=drawColor,anchor=base,inner sep=0pt, outer sep=0pt, scale=  1.00] at ( 88.88, 39.60) {20};

\node[text=drawColor,anchor=base,inner sep=0pt, outer sep=0pt, scale=  1.00] at (121.94, 39.60) {40};

\node[text=drawColor,anchor=base,inner sep=0pt, outer sep=0pt, scale=  1.00] at (155.00, 39.60) {60};

\node[text=drawColor,anchor=base,inner sep=0pt, outer sep=0pt, scale=  1.00] at (188.07, 39.60) {80};

\node[text=drawColor,anchor=base,inner sep=0pt, outer sep=0pt, scale=  1.00] at (221.13, 39.60) {100};

\path[draw=drawColor,line width= 0.4pt,line join=round,line cap=round] ( 49.20, 65.14) -- ( 49.20,163.67);

\path[draw=drawColor,line width= 0.4pt,line join=round,line cap=round] ( 49.20, 65.14) -- ( 43.20, 65.14);

\path[draw=drawColor,line width= 0.4pt,line join=round,line cap=round] ( 49.20, 84.85) -- ( 43.20, 84.85);

\path[draw=drawColor,line width= 0.4pt,line join=round,line cap=round] ( 49.20,104.55) -- ( 43.20,104.55);

\path[draw=drawColor,line width= 0.4pt,line join=round,line cap=round] ( 49.20,124.26) -- ( 43.20,124.26);

\path[draw=drawColor,line width= 0.4pt,line join=round,line cap=round] ( 49.20,143.96) -- ( 43.20,143.96);

\path[draw=drawColor,line width= 0.4pt,line join=round,line cap=round] ( 49.20,163.67) -- ( 43.20,163.67);

\node[text=drawColor,rotate= 90.00,anchor=base,inner sep=0pt, outer sep=0pt, scale=  1.00] at ( 34.80, 65.14) {0};

\node[text=drawColor,rotate= 90.00,anchor=base,inner sep=0pt, outer sep=0pt, scale=  1.00] at ( 34.80, 84.85) {20};

\node[text=drawColor,rotate= 90.00,anchor=base,inner sep=0pt, outer sep=0pt, scale=  1.00] at ( 34.80,104.55) {40};

\node[text=drawColor,rotate= 90.00,anchor=base,inner sep=0pt, outer sep=0pt, scale=  1.00] at ( 34.80,124.26) {60};

\node[text=drawColor,rotate= 90.00,anchor=base,inner sep=0pt, outer sep=0pt, scale=  1.00] at ( 34.80,143.96) {80};

\path[draw=drawColor,line width= 0.4pt,line join=round,line cap=round] ( 49.20, 61.20) --
	(227.75, 61.20) --
	(227.75,167.61) --
	( 49.20,167.61) --
	( 49.20, 61.20);
\end{scope}
\begin{scope}
\path[clip] (  0.00,  0.00) rectangle (252.94,216.81);
\definecolor[named]{drawColor}{rgb}{0.00,0.00,0.00}

\node[text=drawColor,anchor=base,inner sep=0pt, outer sep=0pt, scale=  1.00] at (138.47, 15.60) {Citations (all)};

\node[text=drawColor,rotate= 90.00,anchor=base,inner sep=0pt, outer sep=0pt, scale=  1.00] at ( 10.80,114.41) {\% of papers};
\end{scope}
\end{tikzpicture}

%% file: figures/0-2.all.tex
\begin{tikzpicture}[x=1pt,y=1pt]
\definecolor[named]{fillColor}{rgb}{1.00,1.00,1.00}
\path[use as bounding box,fill=fillColor,fill opacity=0.00] (0,0) rectangle (252.94,216.81);
\begin{scope}
\path[clip] ( 49.20, 61.20) rectangle (227.75,167.61);
\definecolor[named]{drawColor}{rgb}{0.00,0.00,0.00}

\path[draw=drawColor,line width= 0.4pt,line join=round,line cap=round] ( 55.81,130.00) --
	( 55.81,130.16) --
	( 56.68,130.16) --
	( 56.68,130.43) --
	( 57.54,130.43) --
	( 57.54,131.25) --
	( 58.41,131.25) --
	( 58.41,133.26) --
	( 59.27,133.26) --
	( 59.27,135.54) --
	( 60.14,135.54) --
	( 60.14,137.67) --
	( 61.01,137.67) --
	( 61.01,139.84) --
	( 61.87,139.84) --
	( 61.87,141.27) --
	( 62.74,141.27) --
	( 62.74,142.94) --
	( 63.60,142.94) --
	( 63.60,144.77) --
	( 64.47,144.77) --
	( 64.47,146.07) --
	( 65.33,146.07) --
	( 65.33,147.37) --
	( 66.20,147.37) --
	( 66.20,148.64) --
	( 67.06,148.64) --
	( 67.06,149.67) --
	( 67.93,149.67) --
	( 67.93,150.55) --
	( 68.80,150.55) --
	( 68.80,151.45) --
	( 69.66,151.45) --
	( 69.66,152.24) --
	( 70.53,152.24) --
	( 70.53,152.75) --
	( 71.39,152.75) --
	( 71.39,153.38) --
	( 72.26,153.38) --
	( 72.26,154.21) --
	( 73.12,154.21) --
	( 73.12,155.11) --
	( 73.99,155.11) --
	( 73.99,155.48) --
	( 74.85,155.48) --
	( 74.85,155.85) --
	( 75.72,155.85) --
	( 75.72,156.22) --
	( 76.59,156.22) --
	( 76.59,156.62) --
	( 77.45,156.62) --
	( 77.45,157.04) --
	( 78.32,157.04) --
	( 78.32,157.36) --
	( 79.18,157.36) --
	( 79.18,157.57) --
	( 80.05,157.57) --
	( 80.05,157.94) --
	( 80.91,157.94) --
	( 80.91,158.26) --
	( 81.78,158.26) --
	( 81.78,158.61) --
	( 82.64,158.61) --
	( 82.64,158.95) --
	( 83.51,158.95) --
	( 83.51,159.11) --
	( 84.38,159.11) --
	( 84.38,159.30) --
	( 85.24,159.30) --
	( 85.24,159.56) --
	( 86.11,159.56) --
	( 86.11,159.72) --
	( 86.97,159.72) --
	( 86.97,159.96) --
	( 87.84,159.96) --
	( 87.84,160.28) --
	( 88.70,160.28) --
	( 88.70,160.30) --
	( 89.57,160.30) --
	( 89.57,160.43) --
	( 90.43,160.43) --
	( 90.43,160.75) --
	( 91.30,160.75) --
	( 91.30,160.91) --
	( 92.17,160.91) --
	( 92.17,161.02) --
	( 93.03,161.02) --
	( 93.03,161.18) --
	( 93.90,161.18) --
	( 93.90,161.26) --
	( 94.76,161.26) --
	( 94.76,161.36) --
	( 95.63,161.36) --
	( 95.63,161.47) --
	( 96.49,161.47) --
	( 96.49,161.50) --
	( 97.36,161.50) --
	( 97.36,161.55) --
	( 98.22,161.55) --
	( 98.22,161.60) --
	( 99.09,161.60) --
	( 99.09,161.87) --
	( 99.96,161.87) --
	( 99.96,162.00) --
	(100.82,162.00) --
	(100.82,162.10) --
	(101.69,162.10) --
	(101.69,162.16) --
	(102.55,162.16) --
	(102.55,162.29) --
	(103.42,162.29) --
	(103.42,162.37) --
	(105.15,162.37) --
	(105.15,162.42) --
	(106.01,162.42) --
	(106.01,162.48) --
	(106.88,162.48) --
	(106.88,162.53) --
	(108.61,162.53) --
	(108.61,162.61) --
	(109.48,162.61) --
	(109.48,162.69) --
	(110.34,162.69) --
	(110.34,162.71) --
	(111.21,162.71) --
	(111.21,162.74) --
	(112.07,162.74) --
	(112.07,162.82) --
	(112.94,162.82) --
	(112.94,162.85) --
	(113.80,162.85) --
	(113.80,162.90) --
	(114.67,162.90) --
	(114.67,162.93) --
	(116.40,162.93) --
	(116.40,162.95) --
	(117.27,162.95) --
	(117.27,162.98) --
	(119.00,162.98) --
	(119.00,163.01) --
	(120.73,163.01) --
	(120.73,163.03) --
	(121.59,163.03) --
	(121.59,163.06) --
	(125.06,163.06) --
	(125.06,163.09) --
	(125.92,163.09) --
	(125.92,163.11) --
	(132.85,163.11) --
	(132.85,163.14) --
	(133.71,163.14) --
	(133.71,163.19) --
	(135.44,163.19) --
	(135.44,163.24) --
	(137.17,163.24) --
	(137.17,163.27) --
	(145.83,163.27) --
	(145.83,163.30) --
	(147.56,163.30) --
	(147.56,163.32) --
	(149.29,163.32) --
	(149.29,163.35) --
	(151.02,163.35) --
	(151.02,163.38) --
	(151.89,163.38) --
	(151.89,163.43) --
	(152.75,163.43) --
	(152.75,163.46) --
	(156.22,163.46) --
	(156.22,163.51) --
	(159.68,163.51) --
	(159.68,163.54) --
	(163.14,163.54) --
	(163.14,163.56) --
	(169.20,163.56) --
	(169.20,163.59) --
	(170.93,163.59) --
	(170.93,163.62) --
	(194.30,163.62) --
	(194.30,163.64) --
	(213.34,163.64) --
	(213.34,163.67) --
	(221.13,163.67);
\end{scope}
\begin{scope}
\path[clip] (  0.00,  0.00) rectangle (252.94,216.81);
\definecolor[named]{drawColor}{rgb}{0.00,0.00,0.00}

\path[draw=drawColor,line width= 0.4pt,line join=round,line cap=round] ( 55.81, 61.20) -- (185.64, 61.20);

\path[draw=drawColor,line width= 0.4pt,line join=round,line cap=round] ( 55.81, 61.20) -- ( 55.81, 55.20);

\path[draw=drawColor,line width= 0.4pt,line join=round,line cap=round] ( 99.09, 61.20) -- ( 99.09, 55.20);

\path[draw=drawColor,line width= 0.4pt,line join=round,line cap=round] (142.37, 61.20) -- (142.37, 55.20);

\path[draw=drawColor,line width= 0.4pt,line join=round,line cap=round] (185.64, 61.20) -- (185.64, 55.20);

\node[text=drawColor,anchor=base,inner sep=0pt, outer sep=0pt, scale=  1.00] at ( 55.81, 39.60) {0};

\node[text=drawColor,anchor=base,inner sep=0pt, outer sep=0pt, scale=  1.00] at ( 99.09, 39.60) {50};

\node[text=drawColor,anchor=base,inner sep=0pt, outer sep=0pt, scale=  1.00] at (142.37, 39.60) {100};

\node[text=drawColor,anchor=base,inner sep=0pt, outer sep=0pt, scale=  1.00] at (185.64, 39.60) {150};

\path[draw=drawColor,line width= 0.4pt,line join=round,line cap=round] ( 49.20, 65.14) -- ( 49.20,163.67);

\path[draw=drawColor,line width= 0.4pt,line join=round,line cap=round] ( 49.20, 65.14) -- ( 43.20, 65.14);

\path[draw=drawColor,line width= 0.4pt,line join=round,line cap=round] ( 49.20, 84.85) -- ( 43.20, 84.85);

\path[draw=drawColor,line width= 0.4pt,line join=round,line cap=round] ( 49.20,104.55) -- ( 43.20,104.55);

\path[draw=drawColor,line width= 0.4pt,line join=round,line cap=round] ( 49.20,124.26) -- ( 43.20,124.26);

\path[draw=drawColor,line width= 0.4pt,line join=round,line cap=round] ( 49.20,143.96) -- ( 43.20,143.96);

\path[draw=drawColor,line width= 0.4pt,line join=round,line cap=round] ( 49.20,163.67) -- ( 43.20,163.67);

\node[text=drawColor,rotate= 90.00,anchor=base,inner sep=0pt, outer sep=0pt, scale=  1.00] at ( 34.80, 65.14) {0};

\node[text=drawColor,rotate= 90.00,anchor=base,inner sep=0pt, outer sep=0pt, scale=  1.00] at ( 34.80, 84.85) {20};

\node[text=drawColor,rotate= 90.00,anchor=base,inner sep=0pt, outer sep=0pt, scale=  1.00] at ( 34.80,104.55) {40};

\node[text=drawColor,rotate= 90.00,anchor=base,inner sep=0pt, outer sep=0pt, scale=  1.00] at ( 34.80,124.26) {60};

\node[text=drawColor,rotate= 90.00,anchor=base,inner sep=0pt, outer sep=0pt, scale=  1.00] at ( 34.80,143.96) {80};

\path[draw=drawColor,line width= 0.4pt,line join=round,line cap=round] ( 49.20, 61.20) --
	(227.75, 61.20) --
	(227.75,167.61) --
	( 49.20,167.61) --
	( 49.20, 61.20);
\end{scope}
\begin{scope}
\path[clip] (  0.00,  0.00) rectangle (252.94,216.81);
\definecolor[named]{drawColor}{rgb}{0.00,0.00,0.00}

\node[text=drawColor,anchor=base,inner sep=0pt, outer sep=0pt, scale=  1.00] at (138.47, 15.60) {Citations (within 2 years)};

\node[text=drawColor,rotate= 90.00,anchor=base,inner sep=0pt, outer sep=0pt, scale=  1.00] at ( 10.80,114.41) {\% of papers};
\end{scope}
\end{tikzpicture}

%% file: figures/0-2.50.tex
\begin{tikzpicture}[x=1pt,y=1pt]
\definecolor[named]{fillColor}{rgb}{1.00,1.00,1.00}
\path[use as bounding box,fill=fillColor,fill opacity=0.00] (0,0) rectangle (252.94,216.81);
\begin{scope}
\path[clip] ( 49.20, 61.20) rectangle (227.75,167.61);
\definecolor[named]{drawColor}{rgb}{0.00,0.00,0.00}

\path[draw=drawColor,line width= 0.4pt,line join=round,line cap=round] ( 55.81, 96.34) --
    ( 55.81, 96.66) --
	( 59.12, 96.66) --
	( 59.12, 97.19) --
	( 62.43, 97.19) --
	( 62.43, 98.83) --
	( 65.73, 98.83) --
	( 65.73,102.86) --
	( 69.04,102.86) --
	( 69.04,107.42) --
	( 72.34,107.42) --
	( 72.34,111.66) --
	( 75.65,111.66) --
	( 75.65,116.01) --
	( 78.96,116.01) --
	( 78.96,118.87) --
	( 82.26,118.87) --
	( 82.26,122.21) --
	( 85.57,122.21) --
	( 85.57,125.87) --
	( 88.88,125.87) --
	( 88.88,128.47) --
	( 92.18,128.47) --
	( 92.18,131.06) --
	( 95.49,131.06) --
	( 95.49,133.61) --
	( 98.80,133.61) --
	( 98.80,135.68) --
	(102.10,135.68) --
	(102.10,137.43) --
	(105.41,137.43) --
	(105.41,139.23) --
	(108.72,139.23) --
	(108.72,140.82) --
	(112.02,140.82) --
	(112.02,141.83) --
	(115.33,141.83) --
	(115.33,143.10) --
	(118.63,143.10) --
	(118.63,144.74) --
	(121.94,144.74) --
	(121.94,146.55) --
	(125.25,146.55) --
	(125.25,147.29) --
	(128.55,147.29) --
	(128.55,148.03) --
	(131.86,148.03) --
	(131.86,148.77) --
	(135.17,148.77) --
	(135.17,149.57) --
	(138.47,149.57) --
	(138.47,150.42) --
	(141.78,150.42) --
	(141.78,151.05) --
	(145.09,151.05) --
	(145.09,151.48) --
	(148.39,151.48) --
	(148.39,152.22) --
	(151.70,152.22) --
	(151.70,152.85) --
	(155.00,152.85) --
	(155.00,153.54) --
	(158.31,153.54) --
	(158.31,154.23) --
	(161.62,154.23) --
	(161.62,154.55) --
	(164.92,154.55) --
	(164.92,154.92) --
	(168.23,154.92) --
	(168.23,155.45) --
	(171.54,155.45) --
	(171.54,155.77) --
	(174.84,155.77) --
	(174.84,156.25) --
	(178.15,156.25) --
	(178.15,156.88) --
	(181.46,156.88) --
	(181.46,156.94) --
	(184.76,156.94) --
	(184.76,157.20) --
	(188.07,157.20) --
	(188.07,157.84) --
	(191.37,157.84) --
	(191.37,158.16) --
	(194.68,158.16) --
	(194.68,158.37) --
	(197.99,158.37) --
	(197.99,158.69) --
	(201.29,158.69) --
	(201.29,158.84) --
	(204.60,158.84) --
	(204.60,159.06) --
	(207.91,159.06) --
	(207.91,159.27) --
	(211.21,159.27) --
	(211.21,159.32) --
	(214.52,159.32) --
	(214.52,159.43) --
	(217.83,159.43) --
	(217.83,159.53) --
	(221.13,159.53) --
	(221.13,160.06) --
	(224.44,160.06) --
	(224.44,160.33) --
	(227.75,160.33) --
	(227.75,160.54) --
	(231.05,160.54) --
	(231.05,160.65) --
	(234.36,160.65) --
	(234.36,160.91) --
	(237.66,160.91) --
	(237.66,161.07) --
	(244.28,161.07) --
	(244.28,161.18) --
	(247.58,161.18) --
	(247.58,161.28) --
	(250.89,161.28) --
	(250.89,161.39) --
	(252.94,161.39);
\end{scope}
\begin{scope}
\path[clip] (  0.00,  0.00) rectangle (252.94,216.81);
\definecolor[named]{drawColor}{rgb}{0.00,0.00,0.00}

\path[draw=drawColor,line width= 0.4pt,line join=round,line cap=round] ( 55.81, 61.20) -- (221.13, 61.20);

\path[draw=drawColor,line width= 0.4pt,line join=round,line cap=round] ( 55.81, 61.20) -- ( 55.81, 55.20);

\path[draw=drawColor,line width= 0.4pt,line join=round,line cap=round] ( 88.88, 61.20) -- ( 88.88, 55.20);

\path[draw=drawColor,line width= 0.4pt,line join=round,line cap=round] (121.94, 61.20) -- (121.94, 55.20);

\path[draw=drawColor,line width= 0.4pt,line join=round,line cap=round] (155.00, 61.20) -- (155.00, 55.20);

\path[draw=drawColor,line width= 0.4pt,line join=round,line cap=round] (188.07, 61.20) -- (188.07, 55.20);

\path[draw=drawColor,line width= 0.4pt,line join=round,line cap=round] (221.13, 61.20) -- (221.13, 55.20);

\node[text=drawColor,anchor=base,inner sep=0pt, outer sep=0pt, scale=  1.00] at ( 55.81, 39.60) {0};

\node[text=drawColor,anchor=base,inner sep=0pt, outer sep=0pt, scale=  1.00] at ( 88.88, 39.60) {10};

\node[text=drawColor,anchor=base,inner sep=0pt, outer sep=0pt, scale=  1.00] at (121.94, 39.60) {20};

\node[text=drawColor,anchor=base,inner sep=0pt, outer sep=0pt, scale=  1.00] at (155.00, 39.60) {30};

\node[text=drawColor,anchor=base,inner sep=0pt, outer sep=0pt, scale=  1.00] at (188.07, 39.60) {40};

\node[text=drawColor,anchor=base,inner sep=0pt, outer sep=0pt, scale=  1.00] at (221.13, 39.60) {50};

\path[draw=drawColor,line width= 0.4pt,line join=round,line cap=round] ( 49.20, 65.14) -- ( 49.20,163.67);

\path[draw=drawColor,line width= 0.4pt,line join=round,line cap=round] ( 49.20, 65.14) -- ( 43.20, 65.14);

\path[draw=drawColor,line width= 0.4pt,line join=round,line cap=round] ( 49.20, 84.85) -- ( 43.20, 84.85);

\path[draw=drawColor,line width= 0.4pt,line join=round,line cap=round] ( 49.20,104.55) -- ( 43.20,104.55);

\path[draw=drawColor,line width= 0.4pt,line join=round,line cap=round] ( 49.20,124.26) -- ( 43.20,124.26);

\path[draw=drawColor,line width= 0.4pt,line join=round,line cap=round] ( 49.20,143.96) -- ( 43.20,143.96);

\path[draw=drawColor,line width= 0.4pt,line join=round,line cap=round] ( 49.20,163.67) -- ( 43.20,163.67);

\node[text=drawColor,rotate= 90.00,anchor=base,inner sep=0pt, outer sep=0pt, scale=  1.00] at ( 34.80, 65.14) {50};

\node[text=drawColor,rotate= 90.00,anchor=base,inner sep=0pt, outer sep=0pt, scale=  1.00] at ( 34.80, 84.85) {60};

\node[text=drawColor,rotate= 90.00,anchor=base,inner sep=0pt, outer sep=0pt, scale=  1.00] at ( 34.80,104.55) {70};

\node[text=drawColor,rotate= 90.00,anchor=base,inner sep=0pt, outer sep=0pt, scale=  1.00] at ( 34.80,124.26) {80};

\node[text=drawColor,rotate= 90.00,anchor=base,inner sep=0pt, outer sep=0pt, scale=  1.00] at ( 34.80,143.96) {90};

\path[draw=drawColor,line width= 0.4pt,line join=round,line cap=round] ( 49.20, 61.20) --
	(227.75, 61.20) --
	(227.75,167.61) --
	( 49.20,167.61) --
	( 49.20, 61.20);
\end{scope}
\begin{scope}
\path[clip] (  0.00,  0.00) rectangle (252.94,216.81);
\definecolor[named]{drawColor}{rgb}{0.00,0.00,0.00}

\node[text=drawColor,anchor=base,inner sep=0pt, outer sep=0pt, scale=  1.00] at (138.47, 15.60) {Citations};

\node[text=drawColor,rotate= 90.00,anchor=base,inner sep=0pt, outer sep=0pt, scale=  1.00] at ( 10.80,114.41) {\% of papers};
\end{scope}
\end{tikzpicture}

%% file: main.bbl
\begin{thebibliography}{10}

\bibitem{elsevier2012}
Get found--optimize your research articles for search engines.
\newblock
  \url{http://elsevierconnect.com/get-found-optimize-your-research-articles-fo%
r-search-engines}.
\newblock Accessed: 2012-11-08.

\bibitem{beel2010google}
Joeran Beel and Bela Gipp.
\newblock Academic search engine spam and google scholar's resilience against
  it.
\newblock {\em Journal of electronic publishing}, 13(3), 2010.

\bibitem{beel2010}
J{\"o}ran Beel, Bela Gipp, and Erik Wilde.
\newblock Academic search engine optimization (aseo).
\newblock {\em Journal of Scholarly Publishing}, 41(2):176--190, 2010.

\bibitem{bornmann2008}
Lutz Bornmann and Hans-Dieter Daniel.
\newblock What do citation counts measure? a review of studies on citing
  behavior.
\newblock {\em Journal of Documentation}, 64(1):45--80, 2008.

\bibitem{cyril2010}
Labb{\'e} Cyril.
\newblock Ike antkare one of the great stars in the scientific firmament.
\newblock {\em International Society for Scientometrics and Informetrics
  Newsletter}, 6(2):48--52, 2010.

\bibitem{gargouri2010}
Yassine Gargouri, Chawki Hajjem, Vincent Larivi{\`e}re, Yves Gingras, Les Carr,
  Tim Brody, and Stevan Harnad.
\newblock Self-selected or mandated, open access increases citation impact for
  higher quality research.
\newblock {\em Plos one}, 5(10):e13636, 2010.

\bibitem{lawrence2001}
Steve Lawrence.
\newblock Free online availability substantially increases a paper's impact.
\newblock {\em Nature}, 411(6837):521--521, 2001.

\bibitem{lozano2012}
George~A Lozano, Vincent Larivi{\`e}re, and Yves Gingras.
\newblock The weakening relationship between the impact factor and papers'
  citations in the digital age.
\newblock {\em Journal of the American Society for Information Science and
  Technology}, 63(11):2140--2145, 2012.

\bibitem{meyer2009}
Bertrand Meyer, Christine Choppy, J{\o}rgen Staunstrup, and Jan van Leeuwen.
\newblock Viewpoint: Research evaluation for computer science.
\newblock {\em Commun. ACM}, 52(4):31--34, April 2009.

\bibitem{schonfeld2010}
Roger~C Schonfeld, Ross Housewright, and S~Ithaka.
\newblock Faculty survey 2009: Key strategic insights for libraries,
  publishers, and societies.
\newblock 2010.

\end{thebibliography}
